\documentclass[10pt]{article}
\usepackage{amsmath,amsthm,amssymb,bbm}
\usepackage[latin1]{inputenc}
\usepackage[T1]{fontenc}
\usepackage[english]{babel} 
\usepackage[]{csquotes}
\usepackage{hyperref}      
\usepackage{epigraph}
\usepackage[]{graphicx}
\usepackage{microtype} 	
\usepackage{vmargin}		
\usepackage{amssymb}
\usepackage{amsmath}
\usepackage{latexsym}
\usepackage{amsthm}
\usepackage{bbm}
\usepackage{array}
\usepackage{braket}
\usepackage{microtype}
\usepackage{cite}

\setmarginsrb{20mm}{20mm}{20mm}{20mm}{0mm}{0mm}{0mm}{10mm}
\title{\bf \Large Quantum interferences reconstruction with low homodyne detection efficiency}

\author{\normalsize M. Esposito$^{1, *}$, F. Randi$^1$,  K. Titimbo$^{1,2}$, K.~Zimmermann$^{1,2}$, G.~Kourousias$^3$,\\
 \normalsize A. Curri$^3$, R. Floreanini$^2$, F. Parmigiani$^{1,3, 4}$, D. Fausti$^{1,3}$ and F. Benatti$^{1,2, **}$\\
\\
\small ${}^1$Dipartimento di Fisica, Universit\`a di Trieste, 
34127 Trieste, Italy\\
\small ${}^2$Istituto Nazionale di Fisica Nucleare, Sezione di Trieste, 34014 Trieste, Italy\\
\small ${}^3$Sincrotrone Trieste S.C.p.A., 34127 Basovizza, Italy\\
\small ${}^4$ Institute of Physics II, University of Cologne, Germany\\
\\
\small ${}^{**}$  benatti@ts.infn.it, ${}^*$  martina.esposito@elettra.eu  \\
}

\date{\null}
\begin{document}

\maketitle

\begin{abstract}
Standard quantum state reconstruction techniques indicate that a detection efficiency of $0.5$ is an absolute threshold below which quantum interferences cannot be measured. However, alternative statistical techniques suggest that this threshold can be overcome at the price of increasing the statistics  used for the reconstruction. In the following we present numerical experiments proving that quantum interferences can be measured even with a detection efficiency smaller than $0.5$. At the same time we provide a guideline for handling the tomographic reconstruction of quantum states based on homodyne data collected by low efficiency detectors.
\end{abstract}

\section{Introduction}

Homodyne detection is an experimental method that is used to reconstruct quantum states of coherent light by repeatedly measuring a discrete set of field quadratures \cite{Vogel,Smith,Welsch}. 
Usually, a very high detection efficiency and ad-hoc designed apparatuses with low electronic noise are required \cite{Zavatta}. New methods capable of discriminating between different quantum states of light, even with low detection efficiencies, will pave the road to the application of quantum homodyne detection for studying different physical systems embedded in a high noise environment \cite{PP, dabbicco96, grosse14, esposito14}. For this purpose, specific quantum statistical methods, based on minimax and adaptive estimation of the Wigner function, have been developed in \cite{Butucea,Aubry,Lounici}. These approaches allow for the efficient reconstruction of the Wigner function under any noise condition,  at the price of acquiring larger amounts of data. Hence, they overcome the limits of more conventional pattern function quantum tomography \cite{DAriano1,Kiss1,Herzog,DAriano2,Kiss2,Richter,DAriano3,Lvovsky}. The important consequence of this novel statistical approach is that the $0.5$ detection efficiency threshold can be overcome and quantum tomography is still practicable when the signals are measured with appropriate statistics.
The scope of this paper is to report the results of this method tested by performing numerical experiments. Indeed, we consider a linear superposition of two coherent states 
and numerically generate homodyne data according to the corresponding probability distribution distorted by an independent Gaussian noise simulating efficiencies lower than $0.5$.
By properly expanding the set of numerically generated data, we are able to reconstruct the Wigner function of the linear superposition within errors that are compatible with the theoretical bounds. Our results support the theoretical indications that homodyne reconstruction of linear superposition of quantum states is indeed possible also  at efficiencies lower than 0.5.

\section{Wigner function reconstruction}

Let us consider a quantum system with one degree of freedom described by the Hilbert space $L^2(\mathbb{R})$ of square integrable functions $\psi(x)$ over the real line.
The most general states of  such a system are density matrices $\hat{\rho}$, namely convex combinations of projectors $\vert\psi_j\rangle\langle\psi_j\vert$ onto normalised vector 
states  
$$
\hat{\rho}=\sum_{j}\lambda_j\,\vert\psi_j\rangle\langle\psi_j\vert\ ,\qquad\lambda_j\geq 0\ ,\ \sum_j\lambda_j=1\ .
$$
Any density matrix $\hat{\rho}$ can be completely characterised by the associated Wigner 
function $W_\rho(q,p)$ on the phase-space $(q,p)\in\mathbb{R}^2$; namely, by the 
non positive-definite (pseudo) distribution defined by
\begin{equation}
W_\rho(q,p)=\frac{1}{(2\pi)^2}\int_{\mathbb{R}^2}\,{\rm d}u{\rm d}v \,{\rm e}^{i(uq+vp)}\, {\rm Tr}\Big[\hat{\rho}\,{\rm e}^{-i(u\hat{q}+v\hat{p})}\Big]=\frac{1}{2\pi}\int_{\mathbb{R}}{\rm d}u\,{\rm e}^{i\,u\,p}\,\left<q-v/2\vert\hat{\rho}\vert q+v/2\right>\ .
\label{wigner}
\end{equation}
Here $\hat{q}$ and $\hat{p}$ are the position and momentum operators obeying the commutation relations $[\hat{q}\,,\,\hat{p}]=i$, $\hbar=1$, and $\vert q\pm v/2\rangle$ are eigenstates of $\hat q$: $\hat q\vert q\pm v/2\rangle=(q\pm v/2)\vert q\pm v/2\rangle$.
Notice that $W_\rho(q,p)$ is a square integrable function:
\begin{equation}
\label{HS}
2\pi\,\int_{\mathbb{R}^2}{\rm d}q{\rm d}p\,\left|W_\rho(q,p)\right|^2={\rm Tr}\left({\hat{\rho}}^2\right)\leq 1\ .
\end{equation}
Among the advantages of such a representation, is the possibility of expressing the mean value of any operator $\hat{O}$ with respect to a state $\hat{\rho}$ as a pseudo-expectation with respect to $W_\rho(q,p)$ of an associated function $O(q,p)$ over the phase-space, where
\begin{equation}
O(q,p)=\frac{1}{(2\pi)^2}\int_{\mathbb{R}^2}{\rm d}u\, {\rm d}v\, {\rm e}^{-i(uq+vp)}\, {\rm Tr}\Big[\hat{O}\,{\rm e}^{i(u\hat{q}+v\hat{p})}\Big]\ .
\label{wigner1}
\end{equation}
Indeed, by direct inspection one finds
\begin{equation}
2\pi\, \int_{\mathbb{R}^2}{\rm d}u\, {\rm d}v\, W_\rho(q,p)\,O(q,p)={\rm Tr}\Big(\hat{\rho}\,\hat{O}\Big)\ .
\label{wigner2}
\end{equation}
In homodyne detection, a monochromatic signal photon state is mixed with a coherent reference state, a so-called local oscillator, by a $50/50$ beam splitter. The output is collected by two photodiodes and the difference photocurrent is measured. It can be proved that, when the local oscillator is significantly more intense than the signal, the homodyne photocurrent is proportional to the signal quadrature \cite{Ferraro}.
Denoting by $\hat{a}$ and $\hat{a}^{\dagger}$ the single mode annihilation and creation operators associated with the signal, the quadrature operator is defined as
\begin{eqnarray}
    \hat{x}_{\phi}=\frac{ \hat{a} e^{-i \phi} + \hat{a}^{\dagger}e^{i \phi}}{\sqrt{2}}\ ,
		 \label{quad}
\end{eqnarray}
where $\phi$ is the relative phase between signal and local oscillator. The continuum set of quadratures with 
$\phi \in [0, \pi]$ provides a complete characterization of the signal state.
Using the annihilation and creation operators $\hat{a},\hat{a}^\dag$ one constructs position and
momentum-like operators, $\hat{q}=(\hat{a}+\hat {a}^\dag)/\sqrt{2}$ and
$\hat{p}=(\hat{a}-\hat {a}^\dag)/(i\sqrt{2})$. With respect to the latter, the quadrature operator reads:
\begin{eqnarray}
    \hat{x}_{\phi}=\hat{q}\cos\phi+\hat{p}\sin\phi\ .
		 \label{quad1}
\end{eqnarray}
Quadrature operators have continuous spectrum extending over the whole real line, 
$\hat{x}_\phi\vert x\rangle=x\,\vert x\rangle$; given a generic one-mode photon state associated with a density matrix $\hat{\rho}$, its diagonal elements with respect to the (pseudo) eigenvectors 
\begin{equation}
\label{prob}
p_\rho(x,\phi):=\langle x\vert\hat{\rho}\vert x\rangle\ ,
\end{equation}
represent the probability distribution over the quadrature spectrum. 

In homodyne detection experiments the collected data consist of $n$ pairs of quadrature amplitudes and phases $(X_\ell,\Phi_\ell)$: these can be considered as independent, identically distributed stochastic variables.
Given the probability density $p_\rho(x,\phi)$, one could 
reconstruct the Wigner function by substituting the integration with a sum over the pairs for a sufficiently large number of data. However, the measured values $x$ are typically not the eigenvalues of $\hat{x}_\phi$, rather those of 
\begin{equation}
\label{noisequad}
\hat{x}^\eta_\phi=\sqrt{\eta}\hat{x}_\phi+\sqrt{\frac{1-\eta}{2}}y\ ,\ 0\leq\eta\leq1\ ,
\end{equation}
where $y$ is a normally distributed random variable describing the possible noise that may affect the homodyne detection data and $\eta$ parametrizes the detection efficiency that increases from $0$ to $100\%$ with $\eta$ increasing from $0$ to $1$ \cite{Butucea}.
The noise can safely be considered Gaussian and independent from the statistical properties of the quantum state, that is, $y$ can be considered as independent from $\hat{x}_\phi$. Then, as briefly summarised in Appendix A, the Wigner function is reconstructed from a given set of $n$ measured homodyne pairs $(X_\ell,\Phi_\ell)$, $\ell=1,2,\ldots,n$,
by means of an estimator of the form  \cite{Butucea}
\begin{eqnarray}
\label{estimator1}
&&
W^{\eta,r}_{h,n}(q,p)=W^\eta_{h,n}(q,p)\, \chi_r(q,p)\ ,\quad W^\eta_{h,n}(q,p)=\frac{1}{n}\sum_{\ell=1}^n\,K_h^\eta\left([(q,p);\Phi_\ell]-\frac{X_\ell}{\sqrt{\eta}}\right)\ ,\\
\label{estimator2}
&&K_h^\eta\left([(q,p);\Phi_\ell]-\frac{X_\ell}{\sqrt{\eta}}\right)=\int_{-1/h}^{1/h}{\rm d}\xi\,\frac{|\xi|}{4\pi}\,
{\rm e}^{i\xi(q\cos\Phi_\ell+p\sin\Phi_\ell-X_\ell/\sqrt{\eta})}\, {\rm e}^{\gamma\xi^2}\ .
\end{eqnarray} 
This expression is an approximation of the Wigner function in \eqref{noisyWig} by a sum over $n$ homodyne pairs $(X_\ell,\Phi_\ell)$. The parameter $h$ serves to control the divergent factor $\exp(\gamma\xi^2)$, while $r$, through the characteristic function $\chi_r(q,p)$ of a circle $C_r(0)$ of radius $r$ around the origin, restricts the reconstruction to the points $(q,p)$ such that $q^2+p^2\leq r^2$. Both parameters have to be chosen in order to minimise the reconstruction error which is conveniently measured \cite{Aubry} by the $L^2$-distance between the true Wigner function and the reconstructed one, $\|W_\rho-W^{\eta,r}_{h,n}\|_2$. Since such a distance is a function of the data through $W^{\eta,r}_{h,n}$, the $L^2$-norm has to be averaged over different sets, $M$, of quadrature data:
\begin{equation}
\label{error}
\Delta_{h,n}^{\eta,r}(\hat{\rho})=E\left[\|W_\rho-W^{\eta,r}_{h,n}\|_2^2\right] \equiv
E\left[\int_{\mathbb{R}^2}{\rm d}q{\rm d}p\,
\left|W_\rho(q,p)-W^{\eta,r}_{h,n}(q,p)\right|^2\right]\ ,
\end{equation}
where $E$ denotes the average over the $M$ data samples, each sample consisting of $n$ quadrature pairs $(X_\ell,\Phi_\ell)$ corresponding to measured values of $x_{\phi}$ with $\phi \in [0, \pi]$.
In \cite{Aubry}, an optimal dependence of the parameters $r$ and $h$ upon the number of data, $n$, is obtained by minimizing an upper bound to $\Delta_{h,n}^{\eta,r}(\hat{\rho})$. 
\footnote{\label{foot1}
The functional relation between the parameters $h$ and $r$ on $n$ also depends on an auxiliary parameter $\beta>0$. This was introduced in \cite{Aubry} to characterise the localisation properties on $\mathbb{R}^2$ of the Fourier transforms of the Wigner functions of the following class of density matrices addressed in that context: 
$\displaystyle
\label{class}
\mathcal{A}_{\beta,s,L}~=~\left\{\hat{\rho}\,:\, \int_{\mathbb{R}^2}{\rm d}q{\rm d}p\,\left|
F\left[W_\rho\right](q,p)\right|^2\,{\rm e}^{2\beta(w_1^2+w_2^2)^{s/2}}\,\leq\,(2\pi)^2L 
\right\}$.}


\section{Interfering Coherent States}

Homodyne reconstruction is particularly useful to expose quantum interference effects that typically spoil positivity of the Wigner function: it is exactly these effects that are claimed not to be accessible by homodyne reconstruction in presence of efficiency lower than $50\%$, namely when $\eta$ in \eqref{noisequad} is smaller than $1/2$ \cite{DAriano2}. However, in \cite{Butucea} it is theoretically shown that $\eta<1/2$ only requires increasingly larger data sets for achieving small reconstruction errors.
However, this claim was not put to test in those studies as the values of $\eta$ in the considered numerical experiments were close to $1$. 

Instead, we here  consider values $\eta<1/2$ and reconstruct the Wigner function of the following superposition of coherent states
\begin{equation}
\label{Coh1}
\vert\Psi_\alpha\rangle=\frac{\vert\alpha\rangle+\vert-\alpha\rangle}{\sqrt{2\left(1+{\rm e}^{-2|\alpha|^2}\right)}}\ ,
\qquad
\vert\alpha\rangle={\rm e}^{-|\alpha|^2/2}\,{\rm e}^{\alpha\,a^\dag}\ \vert 0\rangle\ ,
\end{equation}
with $\alpha$ any complex number $\alpha_1+i\alpha_2\in\mathbb{C}$.

\begin{figure}[tbhp]
\centering
\includegraphics[width=0.45\textwidth]{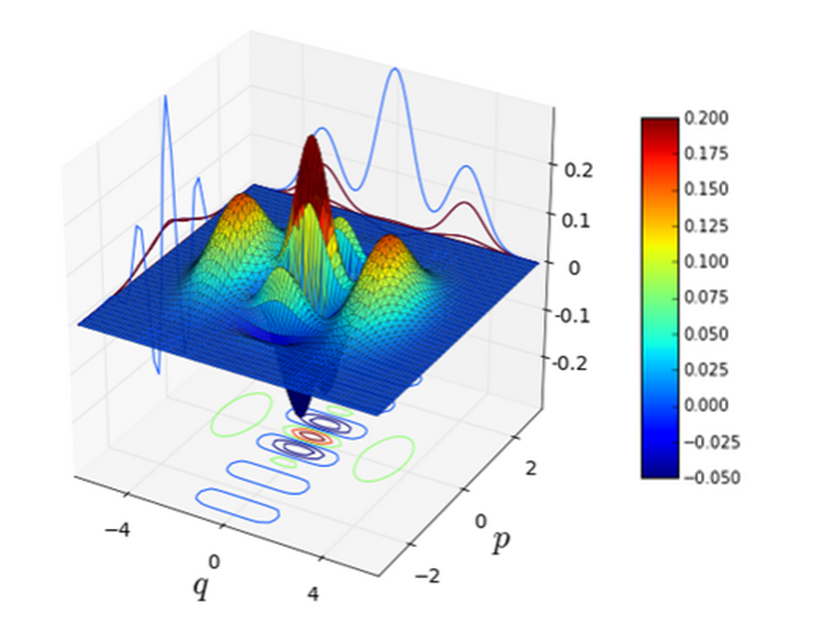}
\caption{Wigner function corresponding to the pure state $\hat{\rho}_\alpha=\vert\Psi_\alpha\rangle\langle\Psi_\alpha\vert$ ($\alpha_1=3/\sqrt{2}$;  $\alpha_2=0$).}
\label{true}
\end{figure}

The Wigner function corresponding to the pure state $\hat{\rho}_\alpha=\vert\Psi_\alpha\rangle\langle\Psi_\alpha\vert$ is shown in Figure \ref{true} for  $\alpha\in\mathbb{R}$. Its general expression together with that of its Fourier transform and of the probability distributions $p_\rho(x,\phi)$ and $p^\eta_\rho(x,\phi)$ are given in Appendix B.

In Appendix C, a derivation is provided of the $L^2$-errors and of the optimal dependence of $h$ and $r$ on the number of data $n$ and on a parameter $\beta$ that takes into account the fast decay of both the Wigner function and its Fourier transform for large values of their arguments.
The following upper bound to the mean square error in \eqref{error} is derived:
\begin{equation}
\label{bound8}
\Delta_{h,n}^{\eta,r}(\hat{\rho}_\alpha)\leq \Delta\ ,\qquad \Delta=\frac{r^2}{n\,h}\,{\rm e}^{2\gamma/h^2}\,\Delta_1(\gamma)\,+\,{\rm e}^{-\beta r^2}\,\Delta_2(\beta)\,+\,{\rm e}^{-\beta/h^2}\,\Delta_3(\beta)\ ,
\end{equation}
with $0<\beta<1/4$ and $\gamma$ as given in \eqref{gamma}.

As explained in Appendix C, the quantities $\Delta_{1,2,3}$ do not depend on $h$, $r$ and $n$. By taking the derivatives with respect to $r$ and $h$, one finds that the upper bound to the mean square deviation is minimised, for large $n$, by choosing
\begin{equation}
\label{optimala}
r=\frac{1}{h}=\sqrt{\frac{\log n}{\beta+2\gamma}}\ . 
\end{equation}\\

\begin{figure}[htbp]
\centering
\includegraphics[width=0.65\textwidth ]{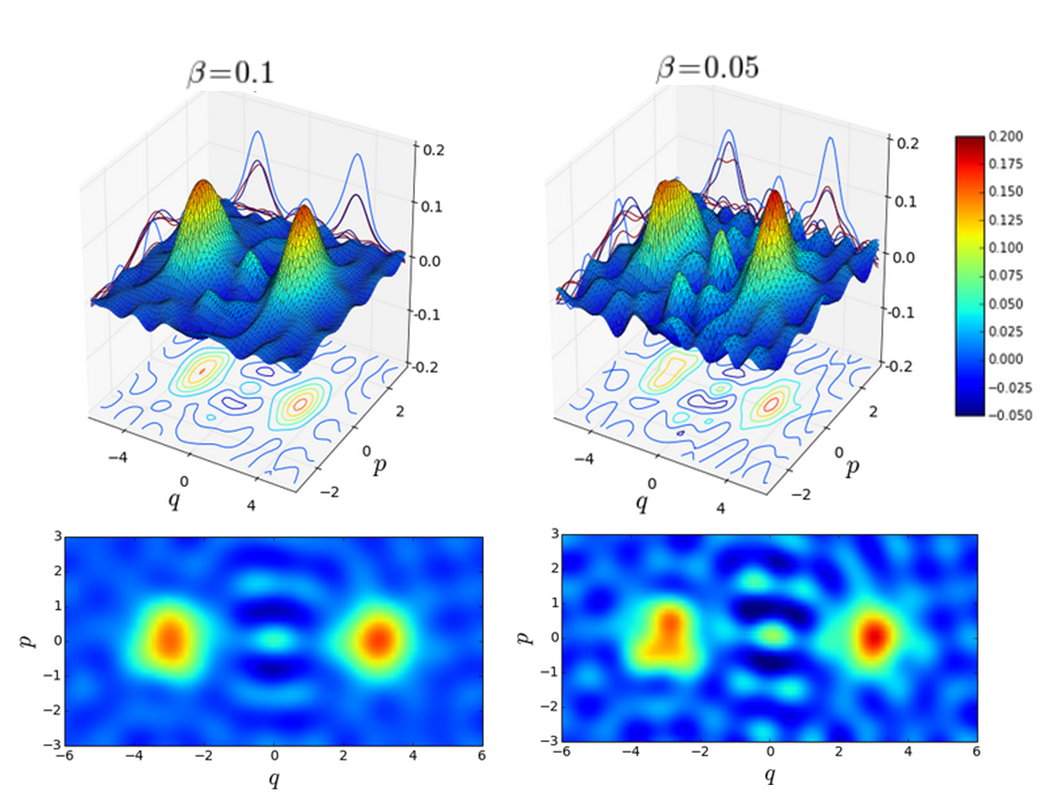}
\caption{Averaged reconstructed Wigner functions $E\left[W^{\eta,r}_{h,n}(q,p)\right]$ over $M=10$ samples of  $n=16\times10^6$ noisy quadrature data (efficiency $\eta = 0.45$). Two different values of $\beta$ are considered.}
\label{045}
\end{figure}

We generated $M=10$ samples of $n=16\times10^6$ quadrature data $(X_\ell,\Phi_\ell)$ distributed according to the noisy probability density $p_\rho^\eta(x,\phi)$ explicitly given in \eqref{aWig4} of Appendix B, considering an efficiency lower than $50\%$ ($\eta=0.45$). Starting from each set of simulated quadrature data we reconstructed the associated Wigner function by means of \eqref{estimator1} and \eqref{estimator2}.
The averaged reconstructed Wigner functions $E\left[W^{\eta,r}_{h,n}(q,p)\right]$ for $\eta = 0.45$ are shown in Figure \ref{045} for two different values of the parameter $\beta$.

The mean square error of the reconstructed Wigner functions has been computed as in \eqref{error} and compared with the mathematically predicted upper bounds $\Delta$. The dependence of the upper bound reconstruction error on the parameter $\beta$ is discussed at the end in Appendix C. In Table \ref{tab2}, we compare the reconstruction errors $\Delta_{h,n}^{\eta,r}(\hat{\rho})$ with their upper bound $\Delta$ for two significant values of $\beta$. 
\begin{table}[hbt]
\begin{center}
\begin{tabular}{|c||c|c|}
	\hline
	$\beta$ & $\Delta_{h,n}^{\eta,r}(\hat{\rho})$ & $\Delta$ \\
	\hline
	\hline
	$0.05$  &   $0.081$  &   $ 2.39 $\\
	\hline
    $0.1$  &   $0.076$  &  $ 26.07 $\\
    \hline
\end{tabular}
	\caption{Calculated $\Delta_{h,n}^{\eta,r}(\hat{\rho})$ for $M=10$ samples of noisy quadrature data ($\eta=0.45$) for two different values of $\beta$. Comparison with the mathematical prediction of the upper bound~$\Delta$.}
	\label{tab2}
\end{center}
\end{table}\\

Despite common belief, the interference features clearly appear in the  reconstructed Wigner function also for efficiencies lower than $50\%$ and the reconstruction errors are compatible with the theoretical predictions. In the next section, we make a quantitative study of the visibility of the interference effects.

\subsection{A witness of interference terms}

The interference effects in the state $\vert\Psi_\alpha\rangle$ can be witnessed by an observable $\hat{O}_\alpha$ of the form
\begin{equation}
\label{observ}
\hat{O}_\alpha=\frac{\vert\alpha\rangle\langle-\alpha\vert\,+\,\vert-\alpha\rangle\langle\alpha\vert}{2\left(1+{\rm e}^{-2|\alpha|^2}\right)}\ .
\end{equation}
With respect to an incoherent mixture of the two coherent states,
\begin{equation}
\label{mixt}
\hat{\rho}_{\alpha\lambda}=\lambda\,\vert\alpha\rangle\langle\alpha\vert\,+\,(1-\lambda)\,\vert-\alpha\rangle\langle-\alpha\vert\ ,\qquad 0\leq\lambda\leq 1\ ,
\end{equation}
its mean value is given by
\begin{equation}
\label{mixmv}
{\rm Tr}\Big(\hat{\rho}_{\alpha\lambda}\,\hat O_\alpha\Big)=\frac{{\rm e}^{-2|\alpha|^2}}{1+{\rm e}^{-2|\alpha|^2}}\ .
\end{equation}
Therefore, from \eqref{wigner1} it follows that the phase-space function $O_\alpha(q,p)$ associated to $\hat O_\alpha$ is 
\begin{equation}
\label{Wignertransf}
O_\alpha(q,p)= \frac{{\rm e}^{-q^2-p^2}\,\cos\left(2\sqrt{2}(q\alpha_2+p\alpha_1)\right)}{\sqrt{\pi}\left(1+{\rm e}^{-2|\alpha|^2}\right)}\ , \qquad \alpha = \alpha_1+i\alpha_2 .
\end{equation}
For details see \eqref{aWig1} and \eqref{aWig3} in Appendix B.

Let us denote by $W_{j,rec}^\alpha(q,p)$, the estimated Wigner function $W^{\eta,r}_{h,n}(q,p)$ in \eqref{estimator1} for the $j$-th set of collected quadrature data. 
It yields a reconstructed mean value 
\begin{equation}
\label{mvrec}
\langle\hat O_\alpha\rangle_{j,rec}=\int_{\mathbb{R}^2}{\rm d}q\,{\rm d}p\, O_\alpha(q,p)\,W^\alpha_{j,rec}(q,p)\ ,
\end{equation}
of which one can compute mean, $\textrm{Av}(<\hat O_\alpha>_{rec})$, and standard deviation, $\textrm{Sd}(<\hat O_\alpha>_{rec})$, with respect to the $M$ sets of collected data:
\begin{eqnarray}
\label{mvrecav}
\textrm{Av}(\langle\hat O_\alpha\rangle_{rec})&=&\frac{1}{M}\sum_{j=1}^M <\hat O_\alpha>_{j, rec}\\ 
\label{mvvar}
\textrm{Sd}(\langle\hat O_\alpha\rangle_{rec})&=&\sqrt{\frac{1}{M}\sum_{j=1}^M\Bigg(\Big(<\hat O_\alpha>_{j,rec}\Big)^2-\Big(\textrm{Av}(<\hat O_\alpha>_{rec}\Big)^2\Bigg)}\ .
\end{eqnarray}

\begin{figure}[htb]
\centering
\includegraphics[width=0.6\textwidth ]{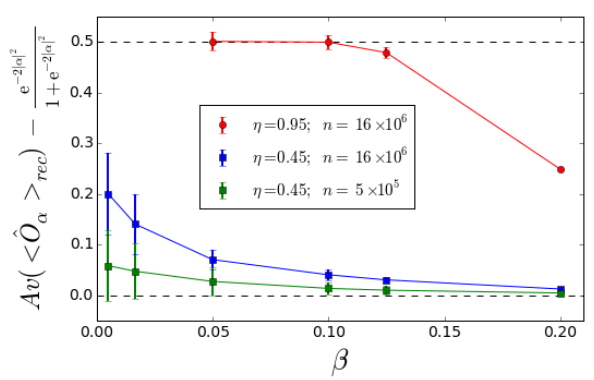}
\caption{$\textrm{Av}(\langle\hat O_\alpha\rangle_{rec})\,-\,\frac{{\rm e}^{-2|\alpha|^2}}{1+{\rm e}^{-2|\alpha|^2}}$ as a function of $\beta$. The error bars represent $\textrm{Sd}~(~\langle~\hat O_\alpha\rangle_{rec}~)$. For each $\beta$, $M=10$ set of $n$  noisy quadrature data have been considered. The square markers refer to $\eta = 0.45$ ($n=16 \times 10^6$ blue marker and $n=5 \times 10^5$ green markers) while the round ones refer to $\eta = 0.95$ ($n=16 \times 10^6$). The error bars for $\eta = 0.95$ have been multiplied by $20$ in order to make them more visible.}
\label{OvsBeta}
\end{figure}

We computed $\textrm{Av}(\langle\hat O_\alpha\rangle_{rec})$ and $\textrm{Sd}(<\hat O_\alpha>_{rec})$ with $M=10$ simulated sets of noisy data with $\eta=0.45$ for two different numbers of simulated quadrature data (see Figure \ref{OvsBeta}). We repeated the procedure for different values of the parameter $\beta$. The results are presented in Figure \ref{OvsBeta}, where the error bars represent the computed $\textrm{Sd}(\langle\hat O_\alpha\rangle_{rec})$. 
\\

In order to be compatible with the interference term present in $\vert\Psi_\alpha\rangle$,
the reconstructed Wigner functions should yield an average incompatible with the incoherent mean value in \eqref{mixmv}, namely such that
\begin{equation}
\label{incompat}
\left|\textrm{Av}(<\hat O_\alpha>_{rec})\,-\,\frac{{\rm e}^{-2|\alpha|^2}}{1+{\rm e}^{-2|\alpha|^2}}\right|\,>\,\textrm{Sd}(<\hat O_\alpha>_{rec})\ .
\end{equation}
We thus see that the condition in \eqref{incompat} is verified for $\eta = 0.45$, that is the reconstructed Wigner functions are not compatible with incoherent superpositions of coherent states, if enough data are considered. 
We also notice that the same behavior is valid for the high efficiency $\eta=0.95$. 

The dependence of the errors on $\beta$ can be understood as follows: when $\beta$ decreases the integration interval in \eqref{estimator2} becomes larger and approaches the exact interval $[-\infty, +\infty]$. Nevertheless this occurs at the price of increasing the reconstruction error. This can be noted both in Figure \ref{OvsBeta} (larger error bars) and in Figure \ref{045} (increasingly noisy effects in the reconstructed Wigner function).  This problem can be overcome with a larger number of data samples $M$, that reduce the reconstruction noise and compensate for the effect of decreasing $\beta$.

\section{Conclusions}

We simulated quadrature data corresponding to high electronic noise and detection efficiencies lower than $0.5$. Under these operating conditions the Wigner function of a linear superposition of two coherent states could be reconstructed using the tomographic techniques developed in \cite{Butucea}. Moreover, by taking into account the decay properties of the Wigner function along with those of its Fourier transform, we have checked that the numerical reconstruction errors are compatible with the theoretical error bounds computed there. 
Furthermore, the reconstruction of the quantum interference pattern of the Wigner function has been supported by a numerical study of the variance of an operatorial interference witness excluding that the reconstructed interferences might be an artefact of the reconstruction algorithm.

We have thus confirmed 1) that, as theoretically predicted in \cite{Aubry,Butucea}, a $0.5$ detection efficiency is not, as often stated in the quantum optics literature, an absolute threshold  below which homodyne quantum state reconstruction is generically impossible and 2) that, instead, by suitably enlarging the size of the set of collected quadrature data, and using alternative techniques different from standard pattern function quantum tomography, one may indeed access quantum features even in low efficiency conditions.

These results also provide the tools for quantum state reconstruction in these operating conditions and set the boundaries of the applicability of the novel statistical approach to homodyne quantum state reconstruction by checking the increase of the number of quadrature data necessary for faithful reconstruction with decreasing detector efficiency.

\section*{Acknowledgments}
 The authors are grateful to Francesca Giusti  for insightful discussions and critical reading of the paper and thank referee number 2 of the manuscript  in reference \cite{esposito14} for stimulating the discussions that led to this work. This work has been supported by a grant from the University of Trieste (FRA 2013).

\appendix
\section{Wigner function reconstruction}

The quadrature probability distribution \eqref{prob} can be conveniently related to the Wigner 
function by passing to polar coordinates
$u=\xi\cos\phi$, $v=\xi\sin\phi$, such that $0\leq\phi\leq\pi$ and $-\infty\leq\xi\leq+\infty$:
\begin{eqnarray}
\nonumber
W_\rho(q,p)&=&\int_0^\pi{\rm d}\phi\int_{-\infty}^{+\infty}{\rm d}\xi\frac{|\xi|}{(2\pi)^2}\,{\rm e}^{i\xi(q\cos\phi+p\sin\phi)}
\, {\rm Tr}\Big[\hat{\rho}\,{\rm e}^{-i\xi(\hat{q}\cos\phi+\hat{p}\sin\phi)}\Big]\\ 
\nonumber
&=&
\int_0^\pi{\rm d}\phi\int_{-\infty}^{+\infty}{\rm d}\xi\frac{|\xi|}{(2\pi)^2}\int_{-\infty}^{+\infty}{\rm d}x\,
{\rm e}^{i\xi(q\cos\phi+p\sin\phi-x)}\,p_\rho(x,\phi)\\
\label{Fourier}
&=&
\int_0^\pi{\rm d}\phi\int_{-\infty}^{+\infty}{\rm d}\xi\frac{|\xi|}{(2\pi)^2}\int_{-\infty}^{+\infty}{\rm d}x\,
{\rm e}^{i\xi(q\cos\phi+p\sin\phi)}\,F[p_\rho(x,\phi)](\xi)\ ,
\end{eqnarray}
where $F[p_\rho(x,\phi)](\xi)$ denotes the Fourier transform with respect to $x$ of the probability distribution:
\begin{equation}
\label{Fourier1}
F[p_\rho(x,\phi)](\xi)=\int_{-\infty}^{+\infty}{\rm d}x\,{\rm e}^{-ix\xi}\,p_\rho(x,\phi)\ .
\end{equation}
Since $y$ can be considered  a normally distributed random variable independent of $\hat{x}_\phi$, the noise affected distribution of the eigenvalues of $\hat{x}_\phi$ in \eqref{noisequad} is given by the following convolution:
\begin{equation}
\label{noisedistr}
p^\eta_\rho(x,\phi)=\int_{-\infty}^{+\infty}{\rm d}u\,\frac{{\rm e}^{-u^2/(1-\eta)}}{\sqrt{\pi(1-\eta)}}\,\frac{p_\rho\Big(\frac{x-u}{\sqrt{\eta}},\phi\Big)}{\sqrt{\eta}}\ .
\end{equation}
Its Fourier transform is connected with that of $p_\rho(x,\phi)$ according to
\begin{equation}
\label{gamma}
F[p_\rho(\,x\,,\phi)](\xi)={\rm e}^{\gamma\xi^2}\,F[p^\eta_\rho(\,x\,,\phi)](\xi/\sqrt{\eta})\ ,\quad\hbox{with}\quad
\gamma:=\frac{1-\eta}{4\eta}\ .
\end{equation}
By inserting $F[p_\rho(\,x\,,\phi)](\xi)$ into \eqref{Fourier1}, one can finally write the Wigner function in terms of the noisy probability distribution $p^\eta_\rho(x,\phi)$:
\begin{equation}
W_\rho(q,p)=\int_0^\pi{\rm d}\phi\int_{-\infty}^{+\infty}{\rm d}\xi\frac{|\xi|}{(2\pi)^2}\int_{-\infty}^{+\infty}{\rm d}x\,
{\rm e}^{i\xi(q\cos\phi+p\sin\phi-x/\sqrt{\eta})}\,{\rm e}^{\gamma\xi^2}\,p_\rho(x,\phi)\ .
\label{noisyWig}
\end{equation}
\section{Coherent state superposition: Wigner function}

The Wigner function corresponding to the pure state $\hat{\rho}_\alpha=\vert\Psi_\alpha\rangle\langle\Psi_\alpha\vert$ and its Fourier transform read 
\begin{eqnarray}
\nonumber
W_\alpha(q,p)&=&\frac{1}{2\pi\left(1+{\rm e}^{-2|\alpha|^2}\right)}\left(
{\rm e}^{-(q-\sqrt{2}\alpha_1)^2-(p-\sqrt{2}\alpha_2)^2}\,+\,
{\rm e}^{-(q+\sqrt{2}\alpha_1)^2-(p+\sqrt{2}\alpha_2)^2}\right.\\
\label{aWig1}
&&\hskip 1cm
\left.+2\,{\rm e}^{-q^2-p^2}\,\cos\left(2\sqrt{2}(q\alpha_2+p\alpha_1)\right)\right) \ ,\\
\nonumber
F[W_\alpha](w_1,w_2)&=&\frac{1}{2\left(1+{\rm e}^{-2|\alpha|^2}\right)}\left(
{\rm e}^{-\frac{(w_1+2\sqrt{2}\alpha_2)^2+(w_2-2\sqrt{2}\alpha_1)^2)}{4}}\,+\,
{\rm e}^{-\frac{(w_1-2\sqrt{2}\alpha_2)^2+(w_2+2\sqrt{2}\alpha_1)^2)}{4}}\right.\\
\label{aWig2}
&&\hskip 1cm
\left.+2\,{\rm e}^{-\frac{w_1^2+w_2^2}{4}}\,\cos\left(\sqrt{2}(w_1\alpha_1+w_2\alpha_2)\right)\right) \ .
\end{eqnarray}
For a generic Wigner function $W_\rho(q,p)$ one computes the quadrature probability density 
$p_\rho(x,\phi)$ in \eqref{prob} by means of the so-called Radon transform:
\begin{equation}
\label{Radon}
\langle x\vert\hat{\rho}\vert x\rangle=\int_{\mathbb{R}}{\rm d}p\,W_\rho(x\cos\phi-p\sin\phi,x\sin\phi+p\cos\phi)\ ,
\end{equation}
It follows that the probability density $p_\rho(x,\phi)$ and the noise-affected  probability density $p^\eta_\rho(x,\phi)$ 
in \eqref{noisedistr} are given by:
\begin{eqnarray}
\nonumber
p_\alpha(x,\phi)&=&\frac{1}{2\sqrt{\pi}\left(1+{\rm e}^{-2|\alpha|^2}\right)}\left(
{\rm e}^{-(x-\sqrt{2}\alpha(\phi))^2}+{\rm e}^{-(x+2\sqrt{2}\alpha(\phi))^2}\right.\\
&&\hskip 1cm\left.
+\,{\rm e}^{-x^2-2\alpha^2(-\phi)}\,2\cos\left(2\sqrt{2}x\beta(-\phi)\right)\right)\ ,
\label{aWig3}\\
\nonumber
p^\eta_\alpha(x,\phi)&=&\frac{1}{2\sqrt{\pi}\left(1+{\rm e}^{-2|\alpha|^2}\right)}\left(
{\rm e}^{-(x-\sqrt{2\eta}\alpha(\phi))^2}+{\rm e}^{-(x-\sqrt{2\eta}\alpha(\phi))^2}\right.\\
&&\hskip 1cm\left.
+2\,
{\rm e}^{-x^2-2|\alpha|^2+2\eta|\beta(-\phi)|^2}\,\cos\left(2\sqrt{2\eta}x\beta(-\phi)\right)\right)\ ,
\label{aWig4}
\end{eqnarray} 
where
$$
\alpha(\phi)=\alpha_1\cos\phi+\alpha_2\sin\phi\ ,\qquad \beta(\phi)=\alpha_2\cos\phi-\alpha_1\sin\phi\ .
$$ 

\section{Upper bound reconstruction error estimation}

Here we derive an upper bound to the mean square error of the reconstructed Wigner function. This analysis is necessary in order to find an optimal functional relation between the free parameters in the reconstruction algorithm such to minimise the reconstruction error. For this purpose we follow the techniques developed in \cite{Aubry} adapting them to the case on a linear superposition of coherent states. Using \eqref{estimator1} and \eqref{estimator2}, one starts by rewriting the error in \eqref{error} as the sum of three contributions:
\begin{eqnarray}
\label{error1}
\Delta_{h,n}^{\eta,r}(\hat{\rho})&=&\int_{C_r(0)}{\rm d}q{\rm d}p\,\left(
E\left[\left|W^{\eta}_{h,n}(q,p)\right|^2\right]-\left|
E\left[W^{\eta}_{h,n}(q,p)\right]\right|^2\right)\\
\label{error2}
&+&\int_{C^c_r(0)}{\rm d}q{\rm d}p\, \left|W_\rho(q,p)\right|^2\  \\
\label{error3}
&+&\int_{C_r(0)}{\rm d}q{\rm d}p\,\left|E\left[W^{\eta}_{h,n}(q,p)\right]-W_\rho(q,p)\right|^2, \,
\end{eqnarray}
$C^c_r(0)$ denoting the region outside the circle $C_r(0)$, where $q^2+p^2>r^2$.
The first and the third term correspond to the variance and bias of the reconstructed Wigner function respectively, while the second term is the error due to restricting the reconstruction to the circle $C_r(0)$. \\
Given a density matrix $\hat{\rho}$, the second term can be directly calculated. This is true also of the bias; indeed, because of the hypothesis that the pairs $(X_\ell,\Phi_\ell)$ are independent identically distributed stochastic variables, it turns out that 
\begin{eqnarray}
\nonumber
E\left[W^{\eta}_{h,n}(q,p)\right]&=&\frac{1}{\pi\,n}\sum_{\ell=1}^n\,E\left[K^\eta_h\left([(q,p);\Phi_\ell]-\frac{X_\ell}{\sqrt{\eta}}\right)\right]=
\frac{1}{\pi}\,E\left[K^\eta_h\left([(q,p);\Phi]-\frac{X}{\sqrt{\eta}}\right)\right]\\
\label{avrec}
&=&\int_0^\pi{\rm d}\phi\int_{-1/h}^{1/h}{\rm d}\xi\frac{|\xi|}{(2\pi)^2}\int_{-\infty}^{+\infty}{\rm d}x\,
{\rm e}^{i\xi(q\cos\phi+p\sin\phi-x/\sqrt{\eta})}\,{\rm e}^{\gamma\xi^2}\,p_\rho(x,\phi)
\end{eqnarray}
differs from the true Wigner function $W_\rho(q,p)$ in \eqref{noisyWig} by the integration over $\xi$ being restricted to the interval $[-1/h,1/h]$. Moreover, its Fourier transform
reads
\begin{eqnarray}
\nonumber
F\left[E\left[W^\eta_{h,n}\right]\right](w)&=&\int_{-\infty}^{+\infty}{\rm d}q\int_{-\infty}^{+\infty}{\rm d}p\,{\rm e}^{-i(qw_1+pw_2)}\,E\left[W^{\eta}_{h,n}(q,p)\right]\\
\label{Favrec}
&=&\chi_{[-1/h,1/h]}(\|w\|)\,F\left[W_\rho\right](w)\ ,\qquad w=(w_1,w_2)\ ,
\end{eqnarray}
where $\chi_{[-1/h,1/h]}(\|w\|)$ is the characteristic function of the interval 
$[-1/h,1/h]$. 
Then, by means of Plancherel equality, 
one gets
\begin{eqnarray}
\nonumber
&&\int_{C_r(0)}{\rm d}q{\rm d}p\,\left|
E\left[W^{\eta}_{h,n}(q,p)\right]-W_\rho(q,p)\right|^2\leq
\int_{\mathbb{R}^2}{\rm d}q{\rm d}p\,\left|
E\left[W^{\eta}_{h,n}(q,p)\right]-W_\rho(q,p)\right|^2\\
\nonumber
&&\hskip 0.5cm=\Big\|E\left[W^\eta_{h,n}\right]-W_\rho\Big\|_2^2=\frac{1}{4\pi^2}
\Big\|F\left[E\left[W^\eta_{h,n}\right]\right]-F\left[W_\rho\right]\Big\|_2^2\\
\label{error4}
&&\hskip 0.5cm =\frac{1}{4\pi^2}\Big\|F\left[W_\rho\right]\,\chi_{[-1/h,1/h]}-F\left[W_\rho\right]\Big\|_2^2=\frac{1}{4\pi^2}\int_{\|w\|\geq1/h}{\rm d}w\,
\Big|F\left[W_\rho\right](w)\Big|^2\ .
\end{eqnarray}

The variance contribution can be estimated as follows: firstly, by using \eqref{estimator1} and \eqref{estimator2}, one recasts it as
\begin{eqnarray}
&&\hskip-0cm\int_{C_r(0)}{\rm d}q{\rm d}p\,\left(
E\left[\left|W^{\eta}_{h,n}(q,p)\right|^2\right]-\left|
E\left[W^{\eta}_{h,n}(q,p)\right]\right|^2\right)=
\nonumber\\
&&\hskip-2.8cm
=\frac{1}{\pi^2\,n}\left\{E\left[\left\|K^{\eta}_{h}\left([\,(q, p)\,;\Phi]-\frac{X}
{\sqrt{\eta}}\right)\,\chi_r(q, p)\right\|^2\right]
\,-\,\left\|E\left[K^{\eta}_{h}\left([\,(q, p)\,;\Phi]-\frac{X}{\sqrt{\eta}}\right)\,\chi_r(q, p)\right]\right\|^2\right\}
\ .
\label{varrec}
\end{eqnarray}
Then, a direct computation of the first contribution yields the upper bound
\begin{equation}
\label{varrec1}
E\left[\left\|K^{\eta}_{h}\left([\,(q, p)\,;\Phi]-\frac{X}
{\sqrt{\eta}}\right)\,\chi_r(q, p)\right\|^2\right]\leq \sqrt{\frac{\pi}{\gamma}}\frac{r^2}{16\,h}\,{\rm e}^{\frac{2\gamma}{h^2}}\,(1+o(1))\ , \ \gamma:=\frac{1-\eta}{4\eta}\ ,
\end{equation}
with $o(1)$ denoting a quantity which vanishes as $h$ when $h\to0$.
On the other hand, the second contribution can be estimated by extending the integration over the whole plane $(q,p)\in\mathbb{R}^2$ and using \eqref{Favrec} together
with \eqref{HS}:
\begin{equation}
\label{varrec2}
\left\|E\left[K^{\eta}_{h}\left([\,(q, p)\,;\Phi]-\frac{X}{\sqrt{\eta}}\right)\,\chi_r(q, p)\right]\right\|^2\leq\frac{1}{4\pi^2}\left\|F\left[W_\rho\right]\right\|^2=
\left\|W_\rho\right\|^2\leq \frac{1}{2\pi}\ .
\end{equation}

Let us consider now the specific case of $\hat{\rho} = \hat{\rho}_{\alpha}$, the superposition of coherent states defined in \eqref{Coh1}.  The auxiliary parameter $\beta$ labelling the class of density matrices $\mathcal{A}_{\beta,s,L}$ in footnote \ref{foot1} with $s=2$ can be used to further optimize the reconstruction error $\Delta_{h, n}^{\eta, r} (\hat{\rho}_\alpha)$. In particular, since $\left|\sum_{j=1}^M z_j\right|^2\leq M\,\sum_{j=1}^M|z_j|^2$, 
we get the upper bounds
\begin{eqnarray}
\label{bound1}
\left|W_\alpha(q,p)\right|&\leq&\frac{2}{\pi}, \qquad 
\left|F\left[W_\alpha\right](w_1,w_2)\right|\leq 2 \\
\nonumber
\left|W_\alpha(q,p)\right|^2&\leq&\frac{3}{4\pi^2}\left(
{\rm e}^{-2(q-\sqrt{2}\alpha_1)^2-2(p-\sqrt{2}\alpha_2)^2}\,+\,
{\rm e}^{-2(q+\sqrt{2}\alpha_1)^2-2(p+\sqrt{2}\alpha_2)^2}\right.\\
\label{bound2}
&&\left.
+4\,{\rm e}^{-2(q^2+p^2)}\right)\leq \frac{3}{2\pi^2}\left({\rm e}^{-(\sqrt{2}R-|\alpha|)^2}+2{\rm e}^{-2R^2}\right)  \ ,\\
\nonumber
\left|F\left[W_\alpha\right](w_1,w_2)\right|^2&\leq& \frac{3}{4}\left(
{\rm e}^{-\frac{(w_1+2\sqrt{2}\alpha_2)^2+(w_2-2\sqrt{2}\alpha_1)^2)}{2}}\,+\,
{\rm e}^{-\frac{(w_1-2\sqrt{2}\alpha_2)^2+(w_2+2\sqrt{2}\alpha_1)^2)}{2}}\right. \\
\label{bound3}
&&\left.
+4\,{\rm e}^{-\frac{w_1^2+w_2^2}{2}}\right)\leq\frac{3}{2}\left({\rm e}^{-(S/\sqrt{2}-2|\alpha|)^2}+2{\rm e}^{-S^2/2}\right)\ ,
\end{eqnarray}
where $R^2=q^2+p^2$ in \eqref{bound2} and $S^2=w_1^2+w_2^2$ in \eqref{bound3} . Then, one derives the upper bounds
\begin{eqnarray}
\label{bound4}
\hskip -1.0cm
\int_{\mathbb{R}^2}{\rm d}q{\rm d}p\left|W_\alpha(q,p)\right|^2\,{\rm e}^{2\beta(q^2+p^2)}
&\leq&\frac{3\left(1+{\rm e}^{4\beta|\alpha|^2/(1-\beta)}\left(1+\frac{2\sqrt{\pi}|\alpha|}{\sqrt{1-\beta}}-{\rm e}^{-4|\alpha|^2/(1-\beta)}\right)\right)}{2\pi(1-\beta)}
\end{eqnarray}
\begin{eqnarray}
\label{bound5}
\nonumber
&&\hskip-1.8cm
\int_{\mathbb{R}^2}{\rm d}w_1{\rm d}w_2\left|F\left[W_\alpha\right](w_1,w_2)\right|^2\,
{\rm e}^{2\beta(w_1^2+w_2^2)}\\
&&\hskip+2.5cm
\leq\frac{6\pi\left(1+{\rm e}^{16\beta|\alpha|^2/(1-4\beta)}\left(1+\frac{2\sqrt{\pi}|\alpha|}{\sqrt{1-4\beta}}-{\rm e}^{-4|\alpha|^2/(1-4\beta)}\right)\right)}{1-4\beta}\ ,
\end{eqnarray}
which simultaneously hold for $0<\beta<1/4$.

Then, by means of the Cauchy-Schwartz inequality, one can estimate the contribution \eqref{error2} to  the error,
\begin{eqnarray}
\nonumber
&&\hskip+1.5cm
\hskip -2cm
\int_{C^c_r(0)}{\rm d}q{\rm d}p\, \left|W_\alpha(q,p)\right|^2 =
\int_{\mathbb{R}^2}{\rm d}q{\rm d}p\, \left|W_\alpha(q,p)\right|^2\,{\rm e}^{\beta(q^2+p^2)}\,{\rm e}^{-\beta(q^2+p^2)}\,\Theta(q^2+p^2-r^2)\\
\nonumber
&\leq&\sqrt{\int_{\mathbb{R}^2}{\rm d}q{\rm d}p\, \left|W_\alpha(q,p)\right|^2\,{\rm e}^{2\beta(q^2+p^2)}}\sqrt{\int_{\mathbb{R}^2}{\rm d}q{\rm d}p\, \left|W_\alpha(q,p)\right|^2\,{\rm e}^{-2\beta(q^2+p^2)}\,\Theta(q^2+p^2-r^2)}\\
&\leq&{\rm e}^{-\beta\,r^2}\,\Delta_2(\beta) \, ,
\label{bound6}
\end{eqnarray}
and similary for \eqref{error4}, 
\begin{eqnarray}
\label{bound7}
&&\hskip0cm
\frac{1}{4\pi^2}\int_{\|w\|\geq1/h}{\rm d}w_1{\rm d}w_2\,
\left|F\left[W_\alpha\right](w_1,w_2)\right|^2\leq{\rm e}^{-\beta/h^2}\,\Delta_3(h)\ ,
\end{eqnarray}
where $\Theta(x)=0$ if $x\leq 0$, $\Theta(x)=1$ otherwise, and
\begin{eqnarray}
\label{bound6a}
\Delta_2(\beta)&=&\sqrt{\frac{3\left(1+{\rm e}^{4\beta|\alpha|^2/(1-\beta)}\left(1+\frac{2\sqrt{\pi}|\alpha|}{\sqrt{1-\beta}}-{\rm e}^{-4|\alpha|^2/(1-\beta)}\right)\right)}{4\pi^2\pi(1-\beta)}}\ ,\\
\Delta_3(\beta)&=&\sqrt{\frac{3\left(1+{\rm e}^{16\beta|\alpha|^2/(1-4\beta)}\left(1+\frac{2\sqrt{\pi}|\alpha|}{\sqrt{1-4\beta}}-{\rm e}^{-4|\alpha|^2/(1-4\beta)}\right)\right)}{4\pi^2\pi(1-4\beta)}}\ .
\label{bound7a}
\end{eqnarray}
Altogether, the previous estimates provide the following upper bound to the mean square error in \eqref{error1}-\eqref{error3}:
\begin{equation}
\label{bound8a}
\Delta_{h,n}^{\eta,r}(\hat{\rho}_\alpha)\leq \Delta\ ,\qquad \Delta=\frac{r^2}{n\,h}\,{\rm e}^{2\gamma/h^2}\,\Delta_1(\gamma)\,+\,{\rm e}^{-\beta r^2}\,\Delta_2(\beta)\,+\,{\rm e}^{-\beta/h^2 }\,\Delta_3(\beta)\ ,
\end{equation}
where $\Delta_{1,2,3}$ do not depend on $h$, $r$ and $n$ and $\Delta_1(\gamma) = \sqrt{\pi}/(16 \pi^2 \sqrt{\gamma})$ is the leading order term in \eqref{varrec1}.
By setting  the derivatives with respect to $r$ and $h$ of the right hand side equal to $0$, one finds 
\begin{eqnarray}
\label{bound9}
\frac{2\gamma}{h^2}+\beta r^2&=&\log n\,+\,\log\left(\beta h\frac{\Delta_2(\beta)}{\Delta_1(\gamma)}\right)\\
\label{bound10}
\frac{2\gamma+\beta}{h^2}&=&\log n\,+\,\log\left(\frac{2\beta h}{r^2(h^2+4\gamma)}\frac{\Delta_3(\beta)}{\Delta_1(\gamma)}\right)\ .
\end{eqnarray}
Whenever $\beta$ is such that the arguments of the logarithms are much smaller than the number of data $n$, to leading order in $n$ the upper bound to the mean square deviation is minimised by 
\begin{equation}
\label{optimal}
r=\frac{1}{h}=\sqrt{\frac{\log n}{\beta+2\gamma}}\ . 
\end{equation}
The range of possible values of $\beta$ is $0\leq\beta\leq1/4$. However, the upper bound $\Delta$ becomes loose when $\beta\to1/4$ and $\beta\to0$.
In the first case, it is the quantity $\Delta_{3}(\beta)$ which diverges, in the second one,  it is the variance contribution which diverges as the logarithm of the number 
of data.
It thus follows that the range of values $\beta\in[\beta_0,\beta_1]$ where the numerical errors $\Delta_{h,n}^{\eta,r}(\hat{\rho}_\alpha)$ are comparable with their upper bounds $\Delta$ is roughly between $\beta_0=0.04$ and $\beta_1=0.10$  for $\eta = 0.45$ as indicated by the following Figure \ref{upperBound}. 
\begin{figure}[hbt]
\centering
\includegraphics[width=0.49\textwidth ]{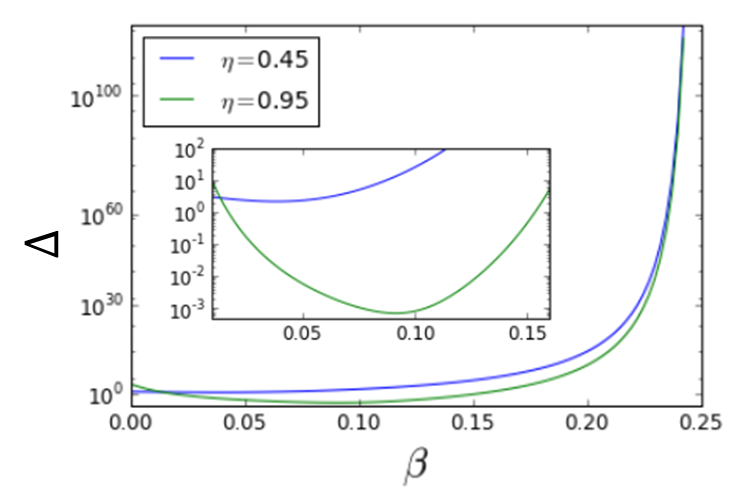}
\caption{Upper bound reconstruction error $\Delta$ as a function of the parameter $\beta$. Two efficiencies $\eta$ are considered.}
\label{upperBound}
\end{figure}

\end{document}